\documentclass[aps,prl,twocolumn,groupedaddress,reprint,showpacs]{revtex4-2}
\usepackage{graphicx}
\usepackage{setspace}
\usepackage{amsmath}
\usepackage{amssymb}
\usepackage{color}
\usepackage{array}
\usepackage{subfigure}
\usepackage{hyperref}
\usepackage{float}
\usepackage{lipsum}

\usepackage[all]{xy}
\newcommand{\RN}[1]{%
  \textup{\uppercase\expandafter{\romannumeral#1}}%
}

\begin{document}

\title{Fast entangling quantum gates with almost-resonant modulated driving}
\author{Xiayang Fan}
\author{Xin Wang}
\thanks{Xiayang Fan and Xin Wang contributed equally to this work.}
\author{Yuan Sun}
\email[email: ]{yuansun@siom.ac.cn}
\affiliation{CAS Key Laboratory of Quantum Optics and Center of Cold Atom Physics, Shanghai Institute of Optics and Fine Mechanics, Chinese Academy of Sciences, Shanghai 201800, China}
\affiliation{University of Chinese Academy of Sciences, Beijing 100049, China}

\begin{abstract}
Recently, the method of off-resonant modulated driving (ORMD) with a special category of synthetic analytical pulses has improved the experimental performance of two- and multi-qubit gates and aroused many interests for further investigations. It particularly offers a helpful tool to the cold atom qubit platform and works well with the Rydberg dipole-dipole interaction. In order to explore more possibilities and wider ranges of options in constructing fast-speed and high-fidelity quantum logic gates, we design and analyze the entangling quantum gates via the almost-resonant modulated driving (ARMD) method. Apart from the apparent distinctions in resonance conditions, the ARMD gate protocols also have its different mechanisms in quantum physics compared with ORMD gate protocols. ARMD gates usually have abrupt phase changes and at certain points during the time evolution. In other words, whilst the modulation forms the key concept of high-fidelity Rydberg blockade gates, the on-off resonance condition can lead to nontrivial nuances in the styles of dynamics. From a more fundamental point of view, the ORMD and the ARMD methods all together belong to the unitary operation family of fast modulated driving with respect to precisely characterized inter-qubit interactions, which usually allows the quantum logic gate to concludes within one continuous pulse. 
\end{abstract}
\maketitle


The two-qubit and multi-qubit quantum logic gates have a vital role in the development of cold atom qubit platform \cite{RevModPhys.82.2313, J.Phys.B.49.202001}. Ever since more than two decades ago, the importance of inter-atomic dipole-dipole interactions has been recognized in the search for gate protocols \cite{PhysRevA.62.052302, PhysRevLett.85.2208}. The Rydberg blockade effect, as a special paradigm, has emerged in experiments as the promising candidate to construct the entangling quantum logic gates of cold atom qubits \cite{nphys1178, nphys1183, PhysRevLett.104.010503}. In the quest for higher fidelity and better connectivity, the family of off-resonant modulated driving (ORMD) Rydberg blockade gates has been theoretically and experimentally established \cite{PhysRevApplied.13.024059, PhysRevA.105.042430, OptEx480513}. The idea of ORMD gates creates a general framework that accommodates further upgrades, especially the suppression of high-frequency components \cite{PhysRevApplied.20.L061002} and the extended Rydberg blockade interaction \cite{Yuan2023arXiv} so far. Together with the progress of single-qubit gates \cite{PhysRevLett.114.100503, PhysRevLett.121.240501, nature604.457, PhysRevX.12.021027}, highly coherent ground-Rydberg transitions \cite{nphys3487, PhysRevLett.104.010502, PhysRevApplied.15.054020, PhysRevLett.129.200501}, scaling up the cold atom array \cite{PhysRevA.92.022336, Weiss2581Science, PhysRevLett.122.203601, Saffman2019prl, Kaufman2020nature, PhysRevApplied.16.034013} and the impressive experimental efforts of ORMD Rydberg blockade gate \cite{PhysRevA.105.042430, Lukin2023nature}, the cold atom qubit platform will find important applications in quantum precision measurement with Rydberg atoms \cite{Browaeys2020review, Browaeys2021nature, Kaufman2022nphys, Browaeys2023nature} besides working on quantum algorithms.

The ORMD gates typically endow the driving lasers with amplitude modulation as the key feature and sometimes include frequency modulation as well, while always operate off-resonantly with respect to the ground-Rydberg transition. The principles can generally be understood as letting the qubit atoms' wave functions receive the correct state-dependent conditional phase shifts via a well-calibrated route in time evolution. With these observations, we start to think about whether we can find an analogy under the on-resonance or close to on-resonance conditions. In other words, it is the time to look into the Rydberg blockade quantum logic gates with almost-resonant modulated driving (ARMD), likely via a single smooth synthetic pulse. In retrospect, interestingly and coincidentally, the early proposals of Rydberg blockade gate use exact on-resonance condition while relying on discrete pulses and often with square wave forms, which turns out to place relatively stringent requirements on the ground-Rydberg coherence \cite{PhysRevApplied.15.054020}. Whilst ORMD and ARMD Rydberg blockades both generate the required conditional phase shifts, the physics of the obtained phases may have drastic differences in their origins. Further investigations along this direction may bring an interesting addition to the quantum geometry \cite{BerryPhasePaper, PhysRevLett.58.1593, PhysRevLett.131.240001}, particularly that the emphasis here manifests itself as the trajectories of time evolution. Potential applications in the atom-photon gate and ensemble qubits also seem interesting for the future \cite{RevModPhys.87.1379, PhysRevLett.115.093601}.

\begin{figure}[h]
\centering
\includegraphics[width=0.432\textwidth]{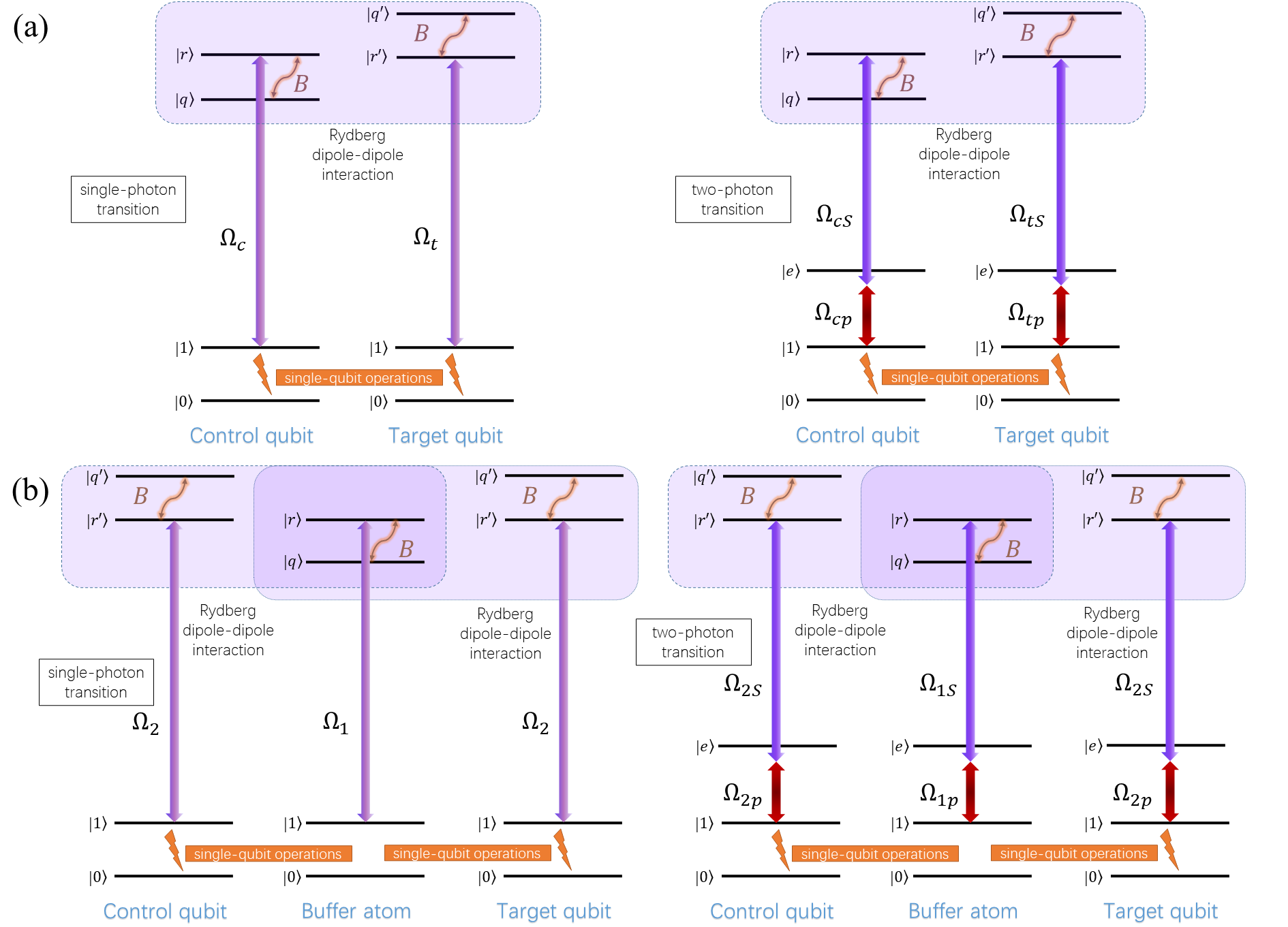}
\includegraphics[width=0.432\textwidth]{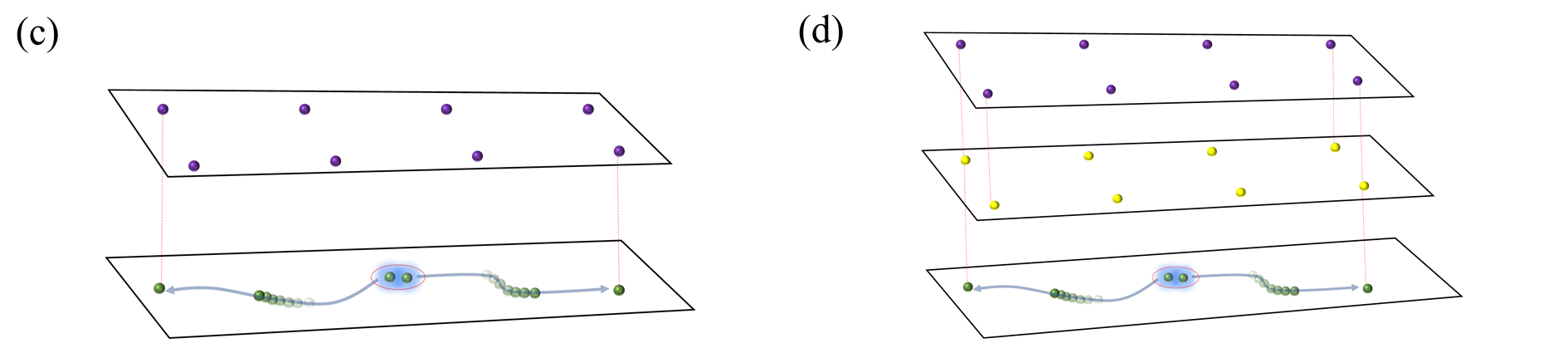}
\caption{(Color online) Typical configurations of single-photon and two-photon ground-Rydberg transitions. (a) The typical two-qubit Rydberg blockade gate. (b) The BAM Rydberg blockade gate with two qubit atoms and one buffer atom. For simplicity, we always denote the qubit register states as $|0\rangle, |1\rangle$, although the qubit atoms can be different species. (c) \& (d) show the concept of obtaining connectivity of FUMOI gates by mechanically moving the messenger atoms, with or without the buffer atom to establish entangling gates between the qubit and the messenger.}
\label{fig1:layout_sketch}
\end{figure}

In this work, we are going to establish and analyze the concept, method and characteristics of the ARMD Rydberg blockade gate and discuss representative examples. Fig. \ref{fig1:layout_sketch} shows the basic ideas and the experimental complexity of ARMD gate approximately stays on the same level as the ORMD gate \cite{PhysRevA.105.042430, Lukin2023nature}. The rest of contents are organized as the following. At first, we propound the two-qubit ARMD gate of the purely two-body system. We will analyze the differences in the physics of ARMD and ORMD gates, and then discuss the broader concept to interpret them. The two- and multi-qubit gate operation can be accomplished within a single synthetic pulse via the unitary operation of fast modulated driving with respect to precisely characterized inter-qubit interactions, or equivalently the fast unitarily modulation observing interactions (FUMOI), with the inter-qubit interaction as the Rydberg dipole-dipole interaction here. Eventually we introduce ARMD in the configuration of buffer-atom-mediated (BAM) two-qubit gate. We evaluate the two-qubit gate fidelity here according to the usual convention \cite{JAMIOLKOWSKI1972275, CHOI1975285, PhysRevA.71.062310, PEDERSEN200747}.

The ARMD method especially suits the two-qubit controlled-Z (CZ) gate. The control and qubit atoms have individualized drivings as they do not even have to be the same element. The on-resonance property naturally appears as the prominent feature of ARMD gate. The overall Hamiltonian is $H_{s1} + H_{s2}$, and $H_{s1}$ corresponds to the single-body process:
\begin{equation}
\label{eq:Hamiltonian_1photon_1body}
H_{s1}/\hbar = \frac{1}{2}\Omega_\text{c}|10\rangle\langle r0| + \frac{1}{2}\Omega_\text{t}|01\rangle\langle 0r| + \text{H.c.},
\end{equation}
with Rabi frequencies $\Omega_\text{c}(t)$ and $\Omega_\text{t}(t)$ coupling $|1\rangle$ and the Rydberg state for the control and target atom, respectively. Meanwhile, the two-body process with idealized Rydberg blockade effect can be described as: 
\begin{equation}
\label{eq:Hamiltonian_1photon_2body}
H_{s2}/\hbar = \frac{1}{2}\Omega_\text{c}|11 \rangle \langle r1| + \frac{1}{2}\Omega_\text{t}|11 \rangle \langle 1r| + \text{H.c.},
\end{equation}
where more details can be found at the supplemental material. 

A wave form $w$ of the Rabi frequency can be expressed with respect to a complete basis $\{g_\nu\}$ for $L^2$ functions on the prescribed time interval: $w=\sum_{\nu=0}^{\infty} \alpha_\nu g_\nu$ with coefficients $\alpha_\nu$'s. Such a framework allows adequate descriptions of possible solutions in the abstract sense, while truncation in the expansion operates conveniently in practice. Aiming at suppressing high-frequencies \cite{PhysRevApplied.20.L061002}, we choose Fourier series as the basis the generate symmetric wave forms. More specifically, the function $f$ of Rabi frequency can be expressed in terms of $[a_0, a_1, \ldots, a_N]$, representing $f(t)=2\pi \times \big(a_0 + \sum_{n=1}^{N} a_n\exp(2\pi i nt/\tau) + a^*_n\exp(-2\pi i nt/\tau) \big)/(2N+1) \text{ MHz}$ for a given reference time $\tau = 0.25\, \mu\text{s}$. Many numerical search algorithms can efficiently find appropriate values of $a_n$'s to satisfy the requirement of quantum logic gates \cite{PhysRevApplied.13.024059, PhysRevApplied.20.L061002}.

\begin{figure}[h]
\centering
\includegraphics[width=0.432\textwidth]{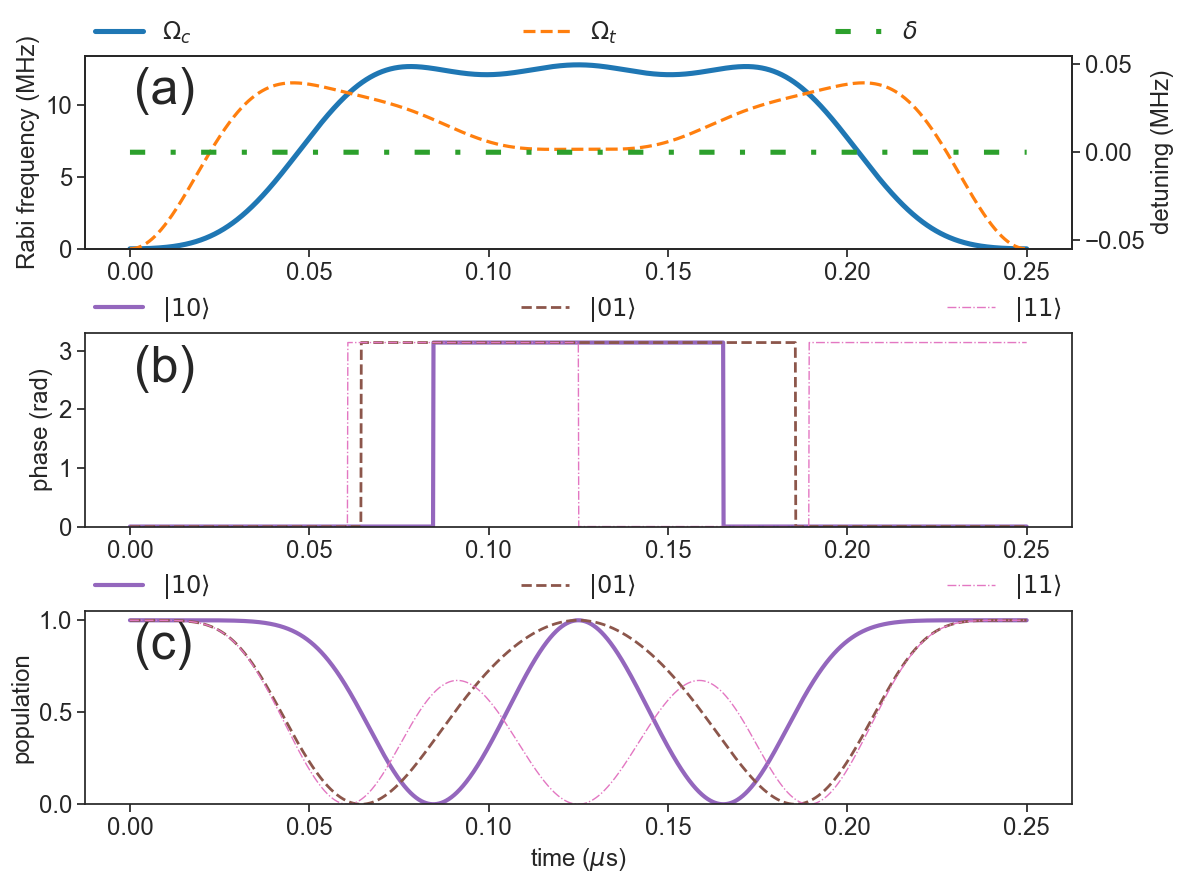}
\caption{(Color online) Sample ARMD gate of one-photon transition for nearly ideal Rydberg blockade. (a) Wave forms of the modulated driving. (b) Phases of wave functions, where the sudden changes represent the phase jump of $\pi$. (c) Populations of wave functions. The calculated gate errors are much less than $10^{-4}$.}
\label{fig:b2gate_1photon}
\end{figure}

We show a typical case in Fig. \ref{fig:b2gate_1photon}, where $\Omega_1(t)$ is given by $[88.01, -36.76, -13.05, 2.07, 4.18, -0.45]$ and $\Omega_2(t)$ is given by $[88.01, -5.93, -20.0, -10.58, -5.0, -2.5]$. Realistically the Rydberg blockade effect deviates from the idealized case, such that the entire system cannot fulfill the perfect on-resonance condition. Moreover, when the detuning term is not exactly zero but close to zero, or even with a small frequency detuning, the ARMD Rydberg blockade gate can still properly function with mostly the same behavior. The accumulated phase in the ARMD gate mostly comes from the dynamical phase while the ORMD gate can have a majority contribution from the geometric phase. Therefore we think it appropriate to use the word of `almost' in the term ARMD.

The two-photon ground-Rydberg transition allows more degrees of freedom from experimental considerations at the cost of extra scattering from the intermediate level. The overall Hamiltonian is $H_\text{c} + H_\text{t} + H_2$, with $H_\text{c}$ describing the single-body process of the control atom:
\begin{equation}
\label{eq:Hamiltonian_2photon_1body}
\frac{H_\text{c}}{\hbar} = \frac{\Omega_\text{cp}}{2}|10\rangle\langle e0| + \frac{\Omega_\text{cS}}{2}|e0\rangle\langle r0| + \text{H.c.} 
+ \Delta |e0\rangle\langle e0|,
\end{equation}
and then the idealized two-body process is:
\begin{align}
\label{eq:Hamiltonian_2photon_2body}
&\frac{H_2}{\hbar} = \frac{\Omega_\text{cp}}{2}|11 \rangle \langle e1| + \frac{\Omega_\text{cS}}{2}|e1 \rangle \langle r1| + \text{H.c.} + \Delta |e1 \rangle \langle r1| \nonumber\\
&\,+\frac{\Omega_\text{tp}}{2}|11 \rangle \langle 1e\rangle + \frac{\Omega_\text{tS}}{2}|1e\rangle\langle 1r| + \text{H.c.} + \Delta |1e \rangle \langle 1r|,
\end{align}
where $\Delta$ is the one-photon detuning and $H_t$ has a similar form but replaces $\Omega_\text{cp}, \Omega_\text{cS}$ with $\Omega_\text{tp}, \Omega_\text{tS}$. 

\begin{figure}[h]
\centering
\includegraphics[width=0.432\textwidth]{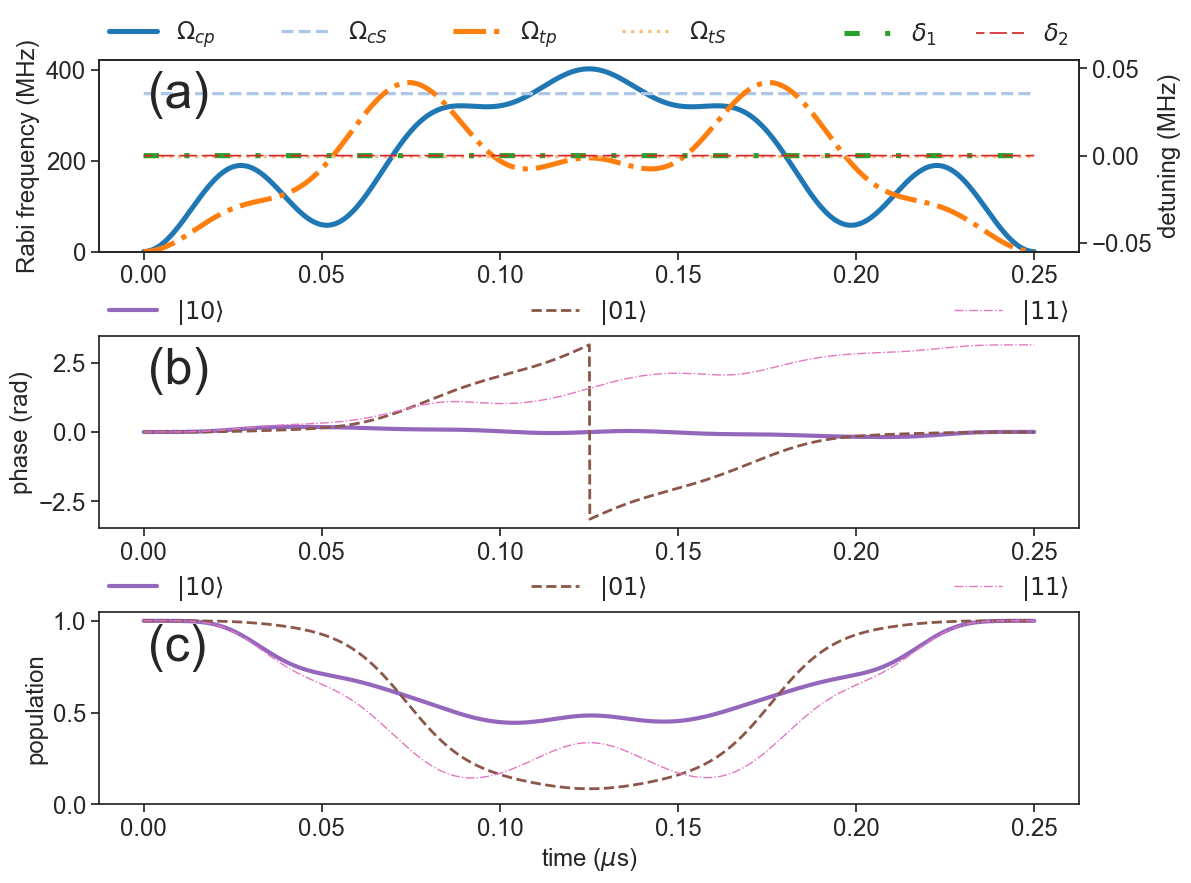}
\caption{(Color online) Sample ARMD gate of two-photon transition with zero two-photon detuning. (a) Wave forms of the modulated driving. (b) Phases of wave functions. (c) Populations of wave functions. The calculated gate errors are less than $10^{-4}$.}
\label{fig:b2gate_2photon_ORMD}
\end{figure}

Naturally we can design a gate which resembles the behavior of by setting the effective two-photon Rabi frequency and two-photon detuning terms to emulate that of one-photon ARMD Rydberg blockade gate. Nevertheless, we discuss a different but interesting case with zero two-photon detuning which appears like the ARMD but actually bearing much of the ORMD gate's property. Such an example is shown in Fig. \ref{fig:b2gate_2photon_ORMD}, where $\Omega_\text{cp}(t)$ is given by [2272.30, -822.50, 210.48, 15.84, -239.97, -300.00], $\Omega_\text{tp}(t)$ is given by [2095.32, -543.21, -560.01, 181.68, 79.91, -206.03], $\Omega_\text{cS} = 2\pi\times 347.79\text{ MHz}$ and $\Omega_\text{tS} = 2\pi\times 208.91\text{ MHz}$.

Together with the previously established ORMD gate \cite{PhysRevApplied.13.024059, PhysRevA.105.042430, OptEx480513, PhysRevApplied.20.L061002, Yuan2023arXiv}, the discussions here about the ARMD gate strongly hint that, while bearing drastic differences, they have deep connections with each other. In fact, they can be regarded as both belonging to a broader concept and we categorize it as the FUMOI gate, which especially works smoothly with Rydberg dipole-dipole interaction in the cold atom qubit platform. With respect to the experimental efforts so far, the relevant efforts have mostly concentrated in the Controlled-PHASE (CPhase) gate. Typically the FUMOI Rydberg blockade gate employs a synthetic smooth modulated pulse that begins and ends at zero value, which generates a unitary transform on the linear space spanned by $|0\rangle, |1\rangle$'s of qubit atoms. Through a fast quantum coherent process of ground-Rydberg atom-light interaction, it conforms to the requirement of two- or multi- qubit quantum logic gate with the help of Rydberg blockade dipole-dipole interaction, which takes place conditionally according the qubits' states. The key ingredient of the FUMOI gate is the carefully tailored modulation process, whose wave forms can be represented by a few discrete numbers such as expansion onto a complete function basis, not only for practical convenience but also without loss of generality. Here, the notion of fast can be understood as much faster than the purely adiabatic process \cite{PhysRevA.90.033408}, or quantitatively speaking, $\bar{E}\tau/\hbar \ll 1$ is significantly violated with $\bar{E}$ as the averaged adiabatic energy difference and $\tau$ as the gate time; often the situation reduces to $\bar{E}\tau/\hbar \sim 1$. 

Therefore, we can now interpret the underlying mechanism of both ARMD and ORMD gates in a unified theme. As the Rabi frequencies gradually increase, the adiabatic states will break the degeneracy in energy. Then the Rabi frequencies and detunings determine the coefficients of initial state of qubit atoms' projection onto these adiabatic states, and this is where ARMD and ORMD starts to differ. For example, equal projection onto the adiabatic states can take place in the ARMD gate while the initial state can mostly overlap with one particular adiabatic state in the ORMD gate. The FUMOI gate induces the conditional phase shifts during the time evolution, resulting in the entangling gate operations. The accumulated phase comes from contributions from both the dynamical phase and the geometric phase. The adiabatic states can cross each other or come relatively close in energy and at such moments the dynamical phase reveals itself in the phase of actual wave function in the qubit register states. Overall, the FUMOI gate contains a large range of possibilities with ARMD and ORMD as two representative generic cases.

\begin{figure}[h]
\centering
\includegraphics[width=0.432\textwidth]{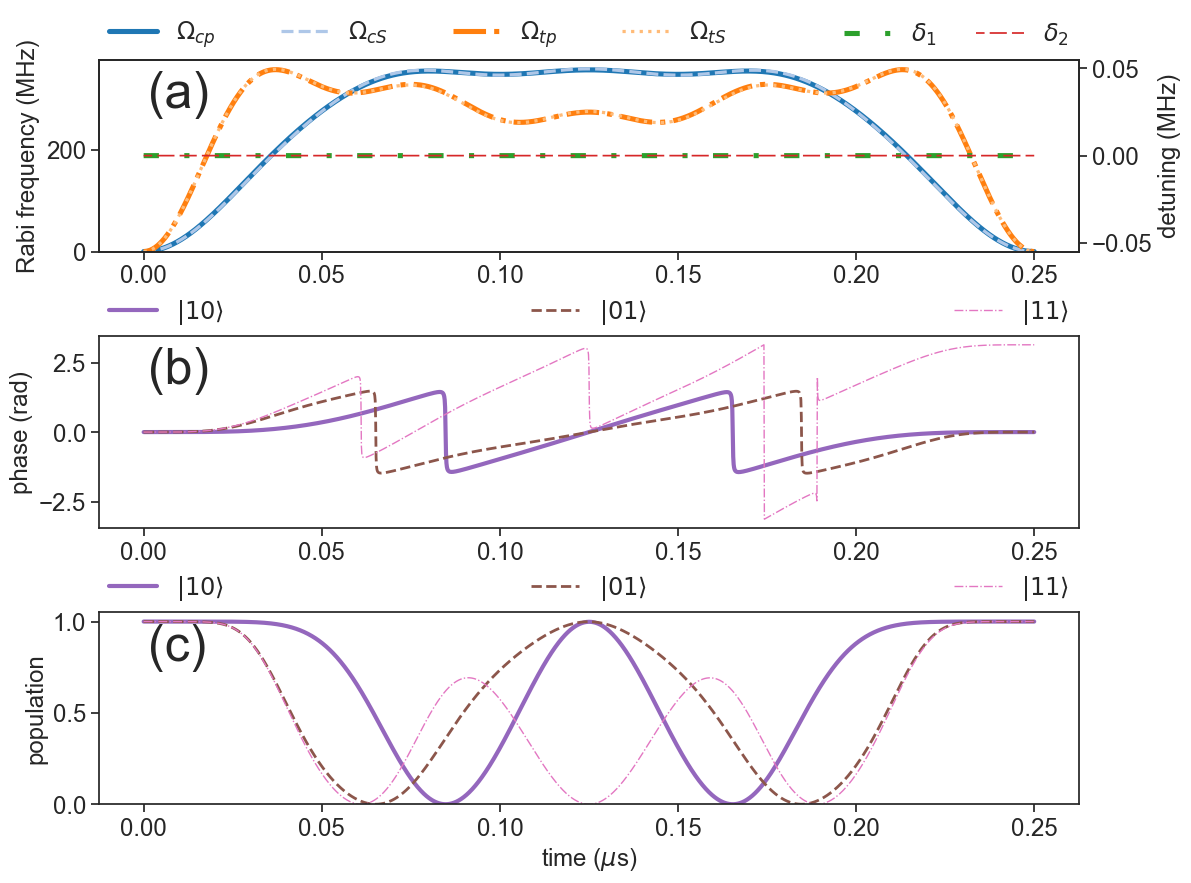}
\caption{(Color online) A sample gate to explain the unified theme of FUMOI gate via two-photon transition. (a) Wave forms of the modulated driving. (b) Phases of wave functions. (c) Populations of wave functions. The calculated gate errors are less than $10^{-4}$.}
\label{fig:b2gate_2photon_nearARMD}
\end{figure}

We provide a concrete example to illustrate the key concepts of the above analysis as shown in Fig. \ref{fig:b2gate_2photon_nearARMD}, thanks to the rich modulation styles provided by the two-photon transition \cite{OptEx480513}. The wave forms are derived by making small variations from that of Fig. \ref{fig:b2gate_1photon}, where $\Omega_\text{cp}(t)$ is given by [2796.08, -867.78, -414.89, -95.59, 1.30, -21.08], $\Omega_\text{tp}(t)$ is given by [2954.90, -249.84, -487.92, -315.42, -234.00, -190.27], $\Omega_\text{cS}(t)$ is given by [2794.86, -872.29, -417.65, -92.25, 10.14, -25.39] and $\Omega_\text{tS}(t)$ is given by [2953.00, -246.82, -485.72, -321.47, -231.64, -190.86]. The contributions from the dynamical phase and geometric phase coexist in the phase accumulation process of this example, and the sudden jumps in the phases of wave functions as shown in Fig. \ref{fig:b2gate_1photon}(b) visualizes the signature of dynamical phase clearly. 

The design of BAM gate \cite{Yuan2023arXiv} aims at suppressing crosstalk and enhancing connectivity, in which an extra buffer atom establishes the linkage between two qubits via nearest-neighbor interaction. The question is then whether the concept of ARMD applies to the BAM gate. Firstly let's look at the case of one-photon transition where one buffer atom mediates two qubit atoms. Trying to better emulate the experimental conditions, the assumption is that the two qubit atoms do not have interactions with each other and the Rydberg dipole-dipole interaction strength between the qubit and buffer atoms take a relatively modest value. Without loss of generality, suppose that the one buffer atom Rabi frequency $\Omega_1$ two qubit atoms Rabi frequency $\Omega_2$ as shown in Fig. \ref{fig1:layout_sketch}(b).

\begin{figure}[h]
\centering
\includegraphics[width=0.432\textwidth]{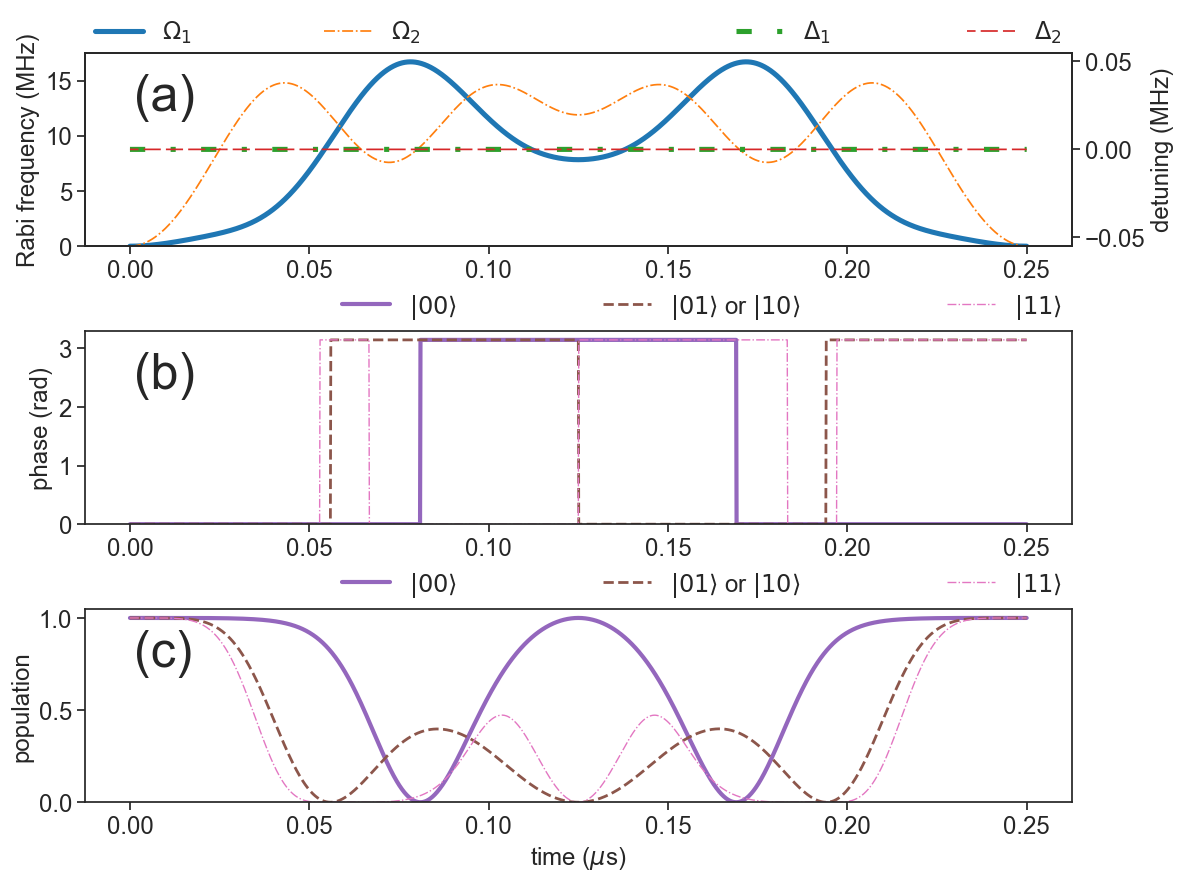}
\caption{(Color online) Sample BAM gate via ARMD of one-photon transition with the Rydberg blockade strength between the buffer and qubit atoms as $B=2\pi\times 50 \text{ MHz}$. (a) The modulated driving. (b) Phases of wave functions. (c) Populations of wave functions. The labeling is with respect to two-qubit basis states, and the calculated gate errors are much less than $10^{-4}$.}
\label{fig:b1gate_1photon}
\end{figure}

Although the transition linkage pattern becomes more complicated than the purely two-body case, it turns out that solutions of ARMD gate exist and are not unique. Fig. \ref{fig:b1gate_1photon} shows a representative result of BAM gate via ARMD with $\Omega_1$ given by [88.00, -33.72, -24.29, 15.71, 1.83, -3.55] and $\Omega_2$ given by [111.82, -19.23, -9.46, -20.0, -13.73, 6.5]. With respect to the BAM gate via two-photon transition, we can also obtain ARMD solutions by choosing parameters to quantitatively emulate that of one-photon transition and omit the discussions here for conciseness. These results indicate that now we can also interpret the BAM gate according to the general concept of FUMOI, and the buffer atom framework can be derived accordingly. For the cold atom qubit platform, the qubit atoms can always keep stationary with the buffer or messenger atoms to establish the necessary connectivity. 

The entangling quantum gate between two remote qubit atoms can be realized via the buffer atom relay or physically moving the messenger atoms. The well-developed experimental technique of mechanically moving and rearranging the atoms \cite{Ahn2016NC} previously only focuses on the qubit atoms, and we can instead keep the qubit atoms immobile but let the messenger atoms relocate. More specifically, the messenger atoms can be prepared in certain entangled states, and then the quantum gate between the messenger and qubit atoms, without or with the buffer atom, will transfer the entanglement into the wanted gate such as the CZ gate between the qubits. Considering scalability, the number of messenger atoms will limit the number of parallel operations of entangling gates but won't limit the number of qubit atoms. For other types of physical qubit platform without the possibility of displacing messengers, the motionless buffer atom relay can help to improve the connectivity \cite{Yuan2023arXiv}. 

According to the analysis of the ARMD Rydberg blockade gate so far, we observe that it naturally fits for constructing the CZ gate. More specifically, the dynamical phase change of ARMD effectively simulates the $\pi$ phase shift of two-level atom model with constant resonant atom-laser interaction. It also has a potentially interesting feature of saving peak laser intensity if designed in an appropriate way.
The possible application of ARMD method in the atom-photon gate \cite{OPTICA.5.001492, PhysRevLett.130.173601} constitutes an interesting question for the next stage. The ARMD method allows full population transfer between the qubit register and Rydberg states, which can potentially take place in the fast readout process of cold atom qubits \cite{PhysRevLett.119.180503, PhysRevLett.119.180504, PRJ_Wang22}. On the other hand, many previously proposed two- and multi- qubit Rydberg blockade gate protocols \cite{PhysRevA.89.030301, PhysRevA.96.042306, PhysRevA.101.022330, PhysRevA.101.062309, PhysRevX.10.021054, PhysRevApplied.16.064031, PhysRevApplied.17.024014, PhysRevApplied.19.044007, ZhangWeiping2023} contain interesting features in the time evolution process, whose comparisons with the ARMD and ORMD methods can possibly offer clues about how to achieve faster operation and better robustness. Another task worthy of investigating is to look for an analogue of this work in other qubit platforms such as the superconducting and nuclear spin qubits. Last but not least, the recent experimental advancement of Rydberg anti-blockade gate \cite{Loh2023NC} will provide another potentially important direction for extending the results of this work.

In conclusion, we have constructed the ARMD Rydberg blockade gate in various forms, including the two-qubit and BAM quantum logic gates. According to our analysis, the ARMD gates and previously established ORMD gates all belong to a unified theme with a much broader concept, namely the FUMOI quantum logic gate. Again, we stress that the FUMOI Rydberg blockade gate applies to qubit atoms of the same or different elements \cite{PhysRevA.92.042710, PhysRevLett.119.160502, PhysRevLett.128.083202, PhysRevX.12.011040}. We have discussed the physics of phase-obtaining process in the FUMOI gate in general, and the various styles draw interesting differences such as the dynamical phase and geometric phase. The contents of this work can straightforwardly extend to multi-qubit Rydberg blockade gates as well as the multi-control or multi-target gates. A lot of future refinements seem promising, especially how to utilize the connections between the different styles of phase accumulation processes and how to further enhance the fidelity against common adverse effects.

\begin{acknowledgments}
This work is supported by the National Natural Science Foundation of China (Grant No. 92165107) and the fundamental research program of the Chinese Academy of Sciences. Xin Wang acknowledges the support from China Postdoctoral Science Foundation (Grant No. 2022M723270). The authors thank Ning Chen, Xiaodong He, Peng Xu and Hui Yan for many discussions.
\end{acknowledgments}

\bibliographystyle{apsrev4-2}
\bibliography{timpani_ref}

\end{document}